\documentclass[twocolumn,amsmath,amssymb]{revtex4}
\usepackage{graphicx}
\usepackage{dcolumn}
\usepackage{bm}

\begin{document}

\title{Device Model for  Graphene Nanoribbon
Phototransistor 
}
\author{Victor~Ryzhii$^{1,4}$, 
Vladimir~Mitin$^{1,2}$,
Maxim~Ryzhii$^{1,4}$, Nadezhda~Ryabova$^{1,4}$, 
and Taiichi~Otsuji$^{3,4}$}
\affiliation{
$^1$Computational Nanoelectronics  Laboratory, University of Aizu, 
Aizu-Wakamatsu, 965-8580, Japan\\
$^2$ Department of Electrical Engineering, University at Buffalo,
Buffalo, NY 14260-1920, USA\\
$^3$ Research Institute for Electrical Communication,
Tohoku University,  Sendai,  980-8577, Japan\\
$^4$Japan Science and Technology Agency, CREST, Tokyo 107-0075, Japan
}

\date{\today}

\begin{abstract}
An analytical device model for a graphene nanoribbon  
phototransistor (GNR-PT) is presented.
GNR-PT is based on an array of graphene nanoribbons 
with the side source and drain contacts, which is 
sandwiched between the highly conducting substrate
and the top gate.
Using the developed model,
we  derive the explicit analytical relationships
for the source-drain current as a function of the intensity
and frequency of the incident
radiation and find the detector responsivity.
It is shown that GNR-PTs can be rather effective photodetectors in
infrared and terahertz ranges of spectrum.
\end{abstract}
%

\maketitle

Photodetectors for far infrared (FIR) and terahertz (THz) ranges
of spectrum are conventionally made of narrow-gap semiconductors
and quantum-well structures. 
In the former case, interband transitions due to the absorption
of photons are used. The operation of the detectors
based on quantum-well structures is associated with the 
electron or hole
intraband (intersubband) transitions~\cite{1,2}.
Some time ago, quantum-dot and quantum-wire detectors were
proposed~\cite{3,4}.
The transition from quantum well structures with two-dimensional
electron (hole) spectrum to low-dimensional structures such as
quantum-wire and quantum-dot structures might lead to
a significant improvement of the FIR and THz 
detectors (see, for instance, ref.~\cite{5} 
and the references therein).  
The utilization of {\it graphene}, i.e.,  
a monolayer of carbon atoms forming a dense honeycomb 
two-dimensional  crystal structure~\cite{6,7,8,9,10} 
opens up tempting prospects in  creation of novel
FIR and THz devices~\cite{11,12,13,14,15},
in particular, novel photodetectors.
One of the most promising metamaterials for FIR and THz detectors
is a patterned graphene which constitutes an array of
graphene nanoribbons (GNRs).
The energy gap between the valence and conduction bands
in GNRs as well as the intraband subbands can be engineered
varying the shape of GNRs, in particular, their width,
which can be defined by lithography.~\cite{16,17,18}.
This opens up the prospects of creation of 
multicolor photodetectors.

In this paper, we propose 
of a graphene nanoribbon phototransistor (GNR-PT)
and evaluate its performance using the developed device model.
The detector proposed has a structure of 
GNR field-effect transistor
consisting of an array of GNRs with the side
source and drain contacts (to each GNR) sandwiched
between the highly conducting substrate and the top gate
electrode. 
The operation of devices with similar structure 
were explored recently 
(see, for instance,~\cite{17,18,19,20,21,22,23}). Here, we study
The structure of GNR-PT under consideration
is schematically shown in Fig.~1.
For the sake of definiteness, we consider a GNR-PT with
optical input 
from the bottom of the structure assuming that the substrate and
the layer sandwiching the GNR array are transparent.

Graphene nanoribbons exhibit the energy spectrum
with a gap between the valence and conduction bands
depending on the nanoribbon width $d$:
\begin{equation}\label{eq1}
\varepsilon_{n}^{\mp}(p) = \pm v\,\sqrt{p^2 + (\pi \hbar/d)^2n^2}.
\end{equation} 
Here $v \simeq 10^8$~cm/s is the characteristic velocity of the electron
(upper sign) and hole (lower sign)
spectra,
$p$ is the momentum  along the nanoribbon, $\hbar$ is the reduced
Planck constant, and $n = 1,2,3,...$ is the subband index.
The quantization  corresponding to Eq.~(1)
of the electron  and hole 
energy spectra
in nanoribbons due to the electron and hole confinement
in one of the lateral directions results in 
the appearance of the band gap $\Delta = 2\pi\hbar/d$ 
between the valence and conduction bands
and in 
a specific
 the density of states (DOS) as a function of the energy.

\begin{figure}
\centerline{\includegraphics[width=8.5cm]{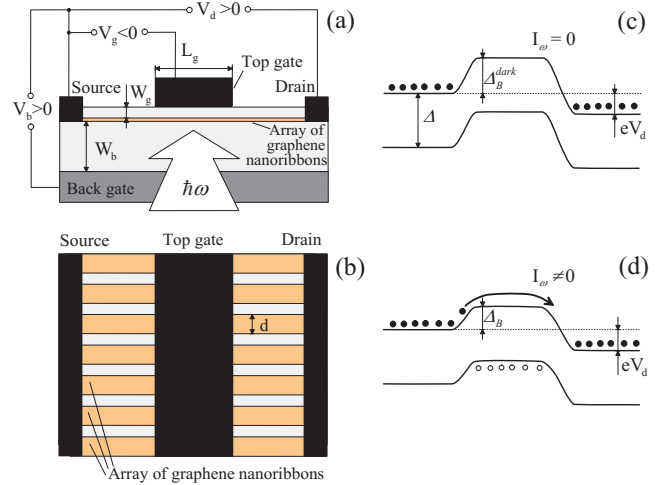}}
\caption{Schematic view of GNR-PT structure side (a)
and top (b)  views, as well as the device 
band diagrams under dark condition (c)
and under irradiation (d). Circles in panels (c) and (d)
corresponds to electrons and holes.}
\end{figure}

The source-drain current along the  GNRs
 associated with the electrons
propagating from the source to the drain and overcoming
the barrier in the center section of the channel (beneath the top gate),
can be presented in the following form:

\begin{equation}\label{eq2}
J = e\Sigma_bv_T\exp\biggl(- \frac{\Delta_B}{k_BT}\biggr)
\biggl[1 -
\exp\biggl(- \frac{eV_d}{k_BT}\biggr)\biggr].
\end{equation}
%
Here, $\Sigma_b = (\ae\,V_b/4\pi\,eW_b)$
is the electron density induced in the channel by the back gate 
voltage $v_T = v\sqrt{4k_BT/\pi\Delta}$
is the electron thermal velocity~\cite{23},
$\Delta_B = - e\varphi_m$ is the height of the barrier in the central
section of the channel,
$\varphi_m < 0$ is the electric potential in its minimum.
Equation~(1) is valid if $V_d \ll V_b$ (but ratio $eV_d/k_BT$ 
can be arbitrary).
The quantities $\varphi_m$ and $\Delta_B$ 
are determined by the gate voltages $V_b$
and $V_g$
and the electron and hole densities in the central section.
The dependence of $\varphi_m$ on the drain voltage $V_d$
is weak provided that $V_d$ is not too large and the length
of the top gate layer $L_g \gg W_b, W_g$. The latter inequality
imply that the ``short-gate'' effects are 
insignificant~\cite{22,23}.
Here $W_b$ and $W_g$ are the thicknesses of the layers
separating the GNR array from the back and top gates, respectively.

In the absence of irradiation,
eq.~(2) leads to the following equation for 
the dark current:~\cite{23}
$$
J^{dark} = v\biggl(\frac{\ae\,V_b}{2\pi^{3/2}W_b}\biggr)
\sqrt{\frac{k_BT}{\Delta}}
$$
\begin{equation}\label{eq3}
\times\exp\biggl(- \frac{\Delta_{B}^{dark}}{k_BT}\biggr)
\biggl[1 -
\exp\biggl(- \frac{eV_d}{k_BT}\biggr)\biggr],
\end{equation}
where $\Delta_B^{dark} = - e[W_bW_g/(W_b + W_g)] (V_b/W_b + V_g/W_g)$
is the height of the barrier in the central section
of the channel under the dark conditions
with $V_b/W_b + V_g/W_g < 0$ in the operation regime, and $\ae$
is the dielectric constant.

Assuming that the photogenerated electrons are quickly swept out
of the central section of the channel by the lateral (along nanoribbons)
electric field, while the photogenerated holes are captured
in this section, the hole density can be found
considering the balance between their photogeneration
and leakage into the source and drain regions.

The equation governing the balance of the photogenerated and thermal
holes
in the central depleted section of the channel,
can be presented as
\begin{equation}\label{eq4}
v_T\Sigma_g \exp\biggl(-\frac{\Delta_B}{k_BT}\biggr)
\biggl[1 + \exp\biggl(- \frac{eV_d}{k_BT}\biggr)\biggr] 
= G_{\omega}L_g,
\end{equation}
where $\Sigma_g$ is the density of holes photogenerated in
the section of the channel beneath the top gate
by the voltage applied to the latter,
$ G_{\omega}$ is the rate of photogeneration
of holes, and $L_g$ is the length of top gate, which is approximately
equal to the length of the depleted section of the channel.
The left-hand side of eq.~(4) is the rate
of the holes leaving the depleted region, whereas the right-hand side
is the net rate of the hole photogeneration in this region.
The effect of holes photogenerated in higly conducting
source and drain sections of the channel is disregarded.

Considering eq.~(4),
and taking into account that
in the gradual channel approximation~\cite{24}
 $\varphi_m \propto \Sigma_g$,
 we arrive at
$$
\Delta_B - \Delta_B^{dark} \simeq - \frac{4\pi\,e^2W}{\ae}\Sigma_g 
$$
\begin{equation}\label{eq5}
\simeq  - \frac{4\pi\,e^2L_gW}{\ae\,v_T}
\frac{\exp(\Delta_B^{dark}/k_BT)}
{[1 + \exp(- eV_dk_BT)]} G_{\omega},
\end{equation}
where $W = W_bW_g/(W_b + W_g)$.
Equation~(5) is valid if the electron density in the section
of the channel under the top gate
is small due to sufficiently strong depletion of this region
by the negatively biased top gate and if the rate of photogeneration
(i.e., the intensity
of radiation) is moderate, so that
 $\Delta_B^{dark} - \Delta_B \ll \Delta_B^{dark}$.

Substituting $\Delta_B$ from eq.~(5) into eq.~(2) and considering
eq.~(3), we arrive at the following equation for
the photocurrent $\Delta J = J - J^{dark}$:

\begin{equation}\label{eq6}
\Delta J \simeq \biggl(\frac{e^2V_b}{k_BT}\biggr)
\biggl[\frac{1 - \exp(-eV_d/k_BT)}
{1 + \exp(-eV_d/k_BT)}\biggr]
\biggl(\frac{W}{W_b}\biggl)L_gG_{\omega}.
\end{equation}

The rate of the photogeneration due to the interband
absorption of incoming radiation is given by
\begin{equation}\label{eq7}
G_{\omega} = \alpha_{\omega} I_{\omega},\qquad
\alpha_{\omega} = \biggl(\frac{4\pi}{c\hbar\omega}\biggr)  Re\,\sigma_{\omega}^{inter},
\end{equation}
where $c$ is the speed of light,
$\hbar\omega$ is the energy of incident  photons,
$I_{\omega}$ is the radiation intensity,
and $Re\,\sigma_{\omega}^{inter}$ is the real part
of the ac interband conductivity.
Considering the energy spectrum of electrons
and holes in nanoribbons, disregarding 
the degeneracy of the electron and hole gases and 
the polarization selectivity
of the interband absorption, and following the procedure
used previously~\cite{12}, this quantity
can be presented as
\begin{equation}\label{eq8}
{\rm Re}\sigma^{inter}(\omega) =\biggl(\frac{e^2}{2\hbar}\biggr)
\sum_{n=1}^{\infty}
\frac{\Delta\cdot\Theta(\hbar\omega - \Delta_n)}{\sqrt{\hbar^2\omega^2 - \Delta_n^2}}.
\end{equation} 
Here, 
$\Delta_n = \varepsilon_{n}^{+}(0) - \varepsilon_{n}^{-}(0) 
= n\,\Delta$, 
and $\Theta(\hbar\omega - \Delta_n)$ is the unity-step function.
Considering eqs.~(7) and (8), 
we obtain
\begin{equation}\label{eq9}
G_{\omega} = 
\sum_{n=1}^{\infty}\frac{\beta\Delta\cdot\Theta(\hbar\omega - n\Delta)}{\sqrt{\hbar^2\omega^2 - n^2\Delta^2}}
\frac{I_{\omega}}{\hbar\omega},
\end{equation} 
where $\beta = 2\pi\, e^2/c\hbar \simeq 2\pi/137 \simeq
4.59\times 10^{-2}$.
The quantity $\beta\Delta/\sqrt{\hbar^2\omega^2 - \Delta^2}$ 
is the effective quantum efficiency
of the GNR array. In the limit $\hbar\omega$ tends to $\Delta$,
the quantum efficiency is limited by 
$\beta\Delta/\Gamma$, where $\Gamma$ is the ``smearing'' of
the valence and conduction band edges due to different imperfections.

Using eqs.~(6) and (9), we obtain the following formula for
the detector responsivity $R = \Delta J/L_gI_{\omega}$:
\begin{figure}[t]
\vspace*{-0.4cm}
\begin{center}
\includegraphics[height=5cm]{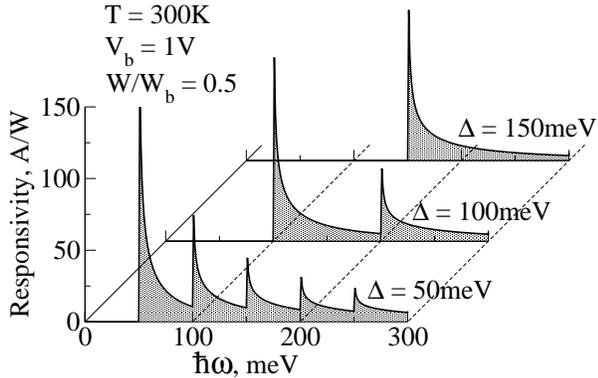}
\caption{Responsivity, $R$, as a function of energy of
incident photons, $\hbar\omega$, 
for GNR-PTs with different band gaps, $\Delta$. 
}
\end{center}
\end{figure}

$$
R \simeq 
\biggl(\frac{W}{W_b}\biggr)\biggl(\frac{eV_b}{k_BT}\biggr)
\biggl[\frac{1 - \exp(-eV_d/k_BT)}
{1 + \exp(-eV_d/k_BT)}\biggr]
$$
\begin{equation}\label{eq10}
\times\sum_{n=1}^{\infty}
\frac{e\beta\Delta\cdot\Theta(\hbar\omega - n\Delta)}
{\hbar\omega\sqrt{\hbar^2\omega^2 - n^2\Delta^2}}.
\end{equation}
It is instructive that the GNR-PT responsivity is independent
of $\Delta_B^{dark}$. This is because 
the effective
life-time of the photogenerated
holes in the deleted section of the channel 
and, therefore, their density increases 
as $\exp(\Delta_B^{dark}/k_BT)$ (see eq.~(4)).
Using eq.~(10)
and setting $W \simeq W_b/2$ and $eV_d \gg k_BT$,
the responsivity maximum (at $\hbar\omega \gtrsim \Delta$) 
and its ratio to the responsivity
minimum (at $\hbar\omega \lesssim 2\Delta$)
can
be estimated, respectively,  as follows: 

\begin{equation}\label{eq11}
{\rm max}\,R \simeq \frac{e\beta}{2\sqrt{2\Gamma\Delta}}\biggl(\frac{eV_b}{k_BT}\biggr) \propto \Delta^{-1/2}T^{-1},
\end{equation}
\begin{equation}\label{eq12}
\frac{{\rm max}\,R}{{\rm min}\,R} \simeq \sqrt{\frac{6\Delta}{\Gamma}}.
\end{equation}
Assuming that $\Delta = 100$~meV,
$\Gamma = 2$~meV, $V_b = 1 - 5$~V, and $T = 300$~K,
we obtain max$R \sim 50 - 250$~A/W.

These values of the responsivity obtained significantly
exceed those for intersubband quantum-well, -wire, and -dot
photodetectors for the IR and THz 
ranges (see, for instance ref.~2).
This is primarily due to higher quantum efficiency and higher
photoelectric gain, which might be  exhibited by GNR-PTs.
The latter is associated with a long life-time of 
the photogenerated holes in the central section of 
the channel because
these holes are confined in this section by relatively
high barriers, so that the photoelectric gain $g \propto
\exp(\Delta_B^{dark}/k_BT) \gg 1$.
The maximum responsivity of GNR-PTs can also exceed
the responsivity of the customary photodetectors
made of narrow gap semiconductors (for example, PbSnTe and CdHgTe),
whose responsivity is about a few A/W~\cite{1,2},
because the former can exhibit rather high quantum efficiency
at the resonances $\hbar\omega = n\,\Delta$ arising
due to the lateral quantization in GNRs.

As follows from eqs.~(11) and (12),  max~$R$ and  min~$R$
markedly increase with decreasing $T$ and $\Delta$.
Figure~2 shows the spectral dependences of the responsivity, $R$,
of GNR-PTs with different energy gaps $\Delta$
(different width of
GNRs) calculated using eq.~(10). It is assumed that  $W/W_b = 0.5$,
$V_b = 1$~V,
and $T = 300$~K.

The principle of the device operation limits
the value $\Delta_B^{dark}$ and, consequently, the photoelectric
gain $g$.
The point is that when the height of the barrier under dark
conditions
$\Delta_B^{dark}$ is comparable
with the band gap $\Delta$, the density of thermal  holes 
can be fairly high. As a result, these holes essentially affect
the potential distribution in the channel that, in turn,
 weakens
the dependence of $\Delta_B$ on the intensity of irradiation.
Taking into account that the Fermi level in the case of devices
with rather large electron densities in the source 
and drain sections
of the channel is close to the bottom of the conduction band,
the pertinent limitation can be presented in the following form:
$\Delta - \Delta_B^{dark} = \Delta +  
eW(V_b/W_b+ V_g/W_g)  \gg k_BT$.
In the above consideration we neglected 
the recombination of the phogenerated holes in
the central section of the channel. The recombination
of the photogenerated holes, primarily, due to
the interband tunneling in the high-field regions
between the quasi-neutral and depleted regions can,
in principle, affect the photoelectric gain. 
This imposes a limitation on the increase of the 
responsivity $R$ with decreasing energy gap $\Delta$ 
corresponding to
eq.~(11).

In conclusion, we developed the GNR-PT  model and
calculated the device characteristics.
It was shown that GNR-PTs can surpass the IR and THz detectors
utilizing other types of quantum structures (in particular,
quantum-well, -wire, and -dot photodetectors).
The GNR-PTs under consideration 
can exhibit substantial technological
advantages, including easier integration with readout circuits,
over the detectors on 
the base of narrow-gap semiconductors
like PbSnTe and CdHgTe.

The work was supported by the Japan Science and 
Technology Agency, CREST,  Japan.
Partial support of the work at UB  by the Air Force Office
of Scientific Research, USA is
also acknowledged.

\end{document}